\documentclass[aps,amsfonts,twocolumn,prd,nofootinbib]{revtex4}
\usepackage{graphicx}

\def\pd      { \partial}
\def\V       {{\mathcal V}}

\def\pd { \partial}
\def\V       {{\mathcal V}}
\def\bra     {{\langle}}
\def\ket     {{\rangle}}
\def\a{\alpha}
\def\b{\beta}

\def\be{\begin{equation}}
\def\ee{\end{equation}}
\newcommand{\eea}{\end{eqnarray}}
\newcommand{\bea}{\begin{eqnarray}}

\def\g{\gamma}
\def\d{\delta}
\def\l{\lambda}
\def\d{\delta}

\def\l{\lambda}

\def\s{\sigma}

\def \p { \phi}
\def \da { \delta \alpha}

\def\pdbk{\bra \phi^\dag \ket}
\def\pbk{\bra \phi \ket}

\def\V2{\mu^2 \pdbk \pbk + \lambda (\pdbk \pbk )^2 }
\usepackage{bm}
\usepackage{epsfig}

\begin{document}

\title{Bubbling the False Vacuum Away}

\author{M.~Gleiser}
\email{gleiser@dartmouth.edu}
\author{B.~Rogers}
\email{rogers@endurance.dartmouth.edu}
\author{J.~Thorarinson}
\email{thorvaldur@dartmouth.edu}

\affiliation{Department of Physics and Astronomy, Dartmouth College,
Hanover, NH  03755, USA\\ \\}

\begin{abstract}
We investigate the role of nonperturbative, bubble-like inhomogeneities on the decay rate of false-vacuum states in two and three-dimensional scalar field theories. The inhomogeneities are induced by setting up large-amplitude oscillations of the field about the false vacuum as, for example, after a rapid quench or in certain models of cosmological inflation. We show that, for a wide range of parameters, the presence of large-amplitude bubble-like inhomogeneities greatly accelerates the decay rate, changing it from the well-known exponential suppression of homogeneous nucleation to a power-law suppression. It is argued that this fast, power-law vacuum decay -- known as resonant nucleation -- is promoted by the presence of long-lived oscillons among the nonperturbative fluctuations about the false vacuum. A phase diagram is obtained distinguishing three possible mechanisms for vacuum decay: homogeneous nucleation, resonant nucleation, and cross-over. Possible applications are briefly discussed.

\end{abstract}

\maketitle

{\section{Introduction}} 

Since the seminal results by Coleman \cite{Coleman}, which can be viewed as an extension of Langer's theory of Homogeneous Nucleation (HN) \cite{Langer} in condensed matter systems to relativistic field theories, there has been a large amount of work dedicated to vacuum decay at both zero and, inspired by Linde's work \cite{Linde}, finite temperatures \cite{decayTfinite}. Several textbooks and review articles describe the topic in detail \cite{decayreviews}.

In high energy physics, the interest in false vacuum decay comes from the fact that models describing the fundamental interactions of matter fields often possess metastable states. For instance, it is still unclear if the quark-hadron transition is or not at least weakly first order \cite{quarkhadron}. Within models of electroweak symmetry breaking, several extensions of the Standard Model, supersymmetric or not, support metastable states \cite{ewtransition}. It is hoped by many that results from the Large Hadron Collider will shed light on this issue, as they may reveal the fundamental mechanism of mass generation. As we move toward the early universe, many models of inflation, including the original old inflation and many others, make use of potentials with metastable states \cite{inflation}. The same is true of grand unified theories. It is thus of great interest to examine under what conditions the predictions from HN theory, which are widely used in the literature, can be trusted. In this work, we will explore the mechanisms by which a false vacuum can decay. As we hope to convince the reader, the effective decay rate is sensitive to the properties of the initial state: the mechanism of false vacuum decay reflects its previous history. Thus, having information about the decay mechanism may help us reconstruct the conditions prevalent at an earlier epoch, when the system under study was still in its metastable state.

This paper is organized as follows: in the next section, we introduce our model and review basic results from false vacuum decay theory. We also briefly review the properties of oscillons, long-lived, time-dependent nonperturbative field configurations, as they play a key role in the present work. We complete the section describing the Hartree effective potential and the details of our $3d$ lattice implementation. In section III, we explore the different mechanisms of vacuum decay, emphasizing the departures from HN. We describe in detail how a power-law decay rate is operative for a wide range of parameters controlling the properties of the initial state and the field interactions. We construct a phase diagram encompassing the three roads toward vacuum decay: HN, resonant nucleation, characterized by power-law decay, and fast cross over. We conclude in section IV, with a summary of our results and possible applications. We include an appendix, describing how to obtain analytically the scaling factor controlling the dependence of the results on lattice spacing.


\section{The Model} 

We are interested in (d+1)-dimensional scalar field theories with conservative dynamics defined by the Hamiltonian

\be 
\label{Hamiltonian}
H[\p] =  \int d^dx \left\{ \frac{1}{2}(\pd_t \phi)^2 + \frac{1}{2} (\pd_i\phi)^2+ V(\phi) \right\}~,  
\ee

\noindent
where the tree-level potential energy density is written for convenience as
\be 
\label{pot}
V(\phi)=\frac{m^2}{2}\phi^2-\frac{\a}{3}\phi^3+\frac{\l}{8}\phi^4~. 
\ee

\begin{figure}
\includegraphics[scale=.6,angle=0]{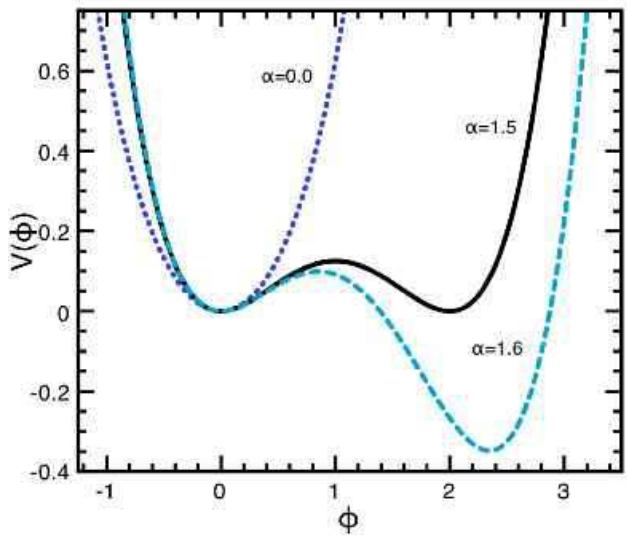}
 \caption{\label{potential} The tree-level potential $V(\p)$ for $\alpha=0.0,1.5,1.6$.  } 
\end{figure}

\noindent
We use units where $\hbar=c=k_B=1$. The parameters $m,~\a$, and $\l$ are positive-definite and temperature independent. With the rescaling, $\p'=\p\sqrt{\l}/m,~x'^{\mu}=x^{\mu}m$, and $\a'=\a/(m\sqrt{\l})$, the potential can be written as $V(\p) = (m^4/\l)V(\p')$, with

\be
\label{respot}
V(\p')=\frac{\p'^2}{2}-\a'\frac{\p'^3}{3}+\frac{\p'^4}{8}~, 
\ee

\noindent
while the Hamiltonian in eq. \ref{Hamiltonian} becomes, $H[\phi] = (m/\l)H[\p']$.
Henceforth we will drop the primes. Note that for $\a=0$ the model is $Z_2$-symmetric, while for $\a=3/2$, $V(\p)$ is a symmetric double-well. For $\a > 3/2$ the minimum at $\p=0$ becomes the false vacuum, and the true vacuum is at $\p_+=\a+\sqrt{\a^2-2}$. In figure  \ref{potential}, we show $V(\p)$ for several values of $\a$.

\subsection{False Vacuum Decay}

Although our results could be extended to models at zero temperature and, thus, with false-vacuum decay controlled solely by quantum tunneling, we will treat here mostly finite-temperature models. We use a finite-temperature description merely for its convenience in setting the properties of the initial state, which we will take to be a state of thermal equilibrium prepared in the potential above with $\a=0$. But before we describe our procedure, it is convenient to briefly review some relevant results from HN.

Consider a model with an asymmetric double-well potential, such as the one of eq. \ref{respot} with $\a>3/2$. The field is initially in thermal equilibrium with only small-amplitude fluctuations about the metastable minimum at $\p_m=0$. One then computes the free energy ${\cal F}=-T\ln {Z}$,
where $Z$ is the partition function, given by the path integral $\int {\cal D}[\p]\exp[-F[\p]/T]$. At finite temperature, $F[\p]$ is the Euclidean free-energy functional of the configuration $\p$, which may contain temperature and quantum corrections to the tree-level theory defined in eq. \ref{Hamiltonian}. To lowest order in perturbation theory, these corrections are included in the effective potential. At $T=0$, one should use the $(d+1)$-dimensional Euclidean effective action $S[\p]$, as opposed to $F[\p]/T$, in the exponent \cite{Linde, decayTfinite}.

The calculation proceeds as follows: first, one identifies the stationary points of $F[\p]$, given by the solutions to the equations of motion. There are two for generic asymmetric double-well potentials, such as the ones in figure \ref{potential}: the false vacuum at $\p_m$, and the spherically-symmetric bounce or critical bubble $\p_b(r)$, found by solving
\be
\label{bounceeq1}
\pd_r^2 \p + \frac{(d-1)}{r} \pd_r \p = \frac{\partial V}{\partial \p}~,
\ee
where $d$ is the number of spatial dimensions. The solution must satisfy the boundary conditions, $\p_b(0) = \p_0$, $\pd_r \p_b(0)=0$, and $\lim_{r\rightarrow \infty} \p(r)=\p_m$. Second, adopting the semi-classical approximation where the path integral is dominated by small fluctuations about the classical paths, one sums over {\it quadratic} fluctuations about the stationary points of the free energy using the saddle-point approximation to the Gaussian path integral. The calculation is completed by summing over the contribution from many bubbles, which makes use of the ``dilute gas'' approximation, where the nucleating bubbles are assumed to be far enough away from each other so as not to overlap, being thus treated as independent events. The end result, the probability per volume per unit time of nucleating a bubble of ``true'' vacuum within the false vacuum background for a system at temperature $T$ is,
\be
\label{decayrate}
\Gamma \sim T^{(d+1)} \exp[-F[\p_b]/T]~.
\ee
The quotes are a reminder that the nucleating bubbles are not initially at the true vacuum, only close to it ($\p_0\neq \p_+$). The sums over bubbles and over quadratic fluctuations about the stationary points give rise to a prefactor which can be roughly approximated by $T^{(d+1)}$ or, in quantum tunneling, by $M^{d+1}$, where $M$ is the relevant mass scale. Here, we will be mainly interested in the exponential factor which, in most scenarios, dominates the false vacuum decay rate. Summarizing, the calculation of false vacuum decay in HN relies on two key assumptions: small (quadratic) fluctuations about the false vacuum and about the critical bubble, and the dilute gas approximation.

\subsection{Thin Wall Approximation}

Although the equation describing the critical bubble, eq. \ref{bounceeq1}, does not, in general, have exact solutions, an estimate of the energy of a critical bubble can be obtained in the ``thin wall'' approximation \cite{Coleman, decayreviews}. Essentially, when the potential $V(\p)$ is nearly degenerate, the critical bubble will have a spatial extension, or ``radius,'' $(R_b)$ much larger than the thickness of the wall separating its interior from its exterior, defined as the region where $\p$ changes abruptly from $\p_0\simeq \p_+$ at $r=0$ to $\p_m$ at $r \rightarrow \infty$. There are thus two main contributions to the free energy (or Euclidean action) of the critical bubble: the wall ($r\sim R_b)$, and the interior of the bubble ($r<R_b$), where one writes $\p_b(r)= \p_+$. The approximate expression for the bubble energy is,

\be E_{\rm tw}\simeq C_d R^{d-1} \left\{ \s - \frac{R}{d} |\Delta V| \right\}, \ee

\noindent
where $|\Delta V|=|V(\p_+)-V(\p_m)|$, and $C_d=2\pi^{d/2}/\Gamma(d/2)$ is related to the volume of a $d$-dimensional sphere of radius $R$ by $V_d=C_d {R^d \over d}$. Also, $\s=\int dr\left [ \frac{1}{2}(\phi'_b)^2 + V(\phi_b)\right ] $ is the surface tension. 
Minimizing with respect to $R$ gives:

\be 
\label{twr}
R_b=\frac{\s (d-1)}{|\Delta V|}, 
\ee
and the nucleation energy-barrier  in the thin wall approximation is,

\be 
\label{twe}
E_{\rm tw}=C_d\frac{(d-1)^{d-1}}{d}\frac{\sigma^{d}}{|\Delta V|^{d-1}}. 
\ee

Within the thin-wall approximation, the bounce solution may be approximately parameterized by $\phi(r)\simeq\frac{\phi_+}{2}(1-\tanh({1\over 2} (r-R_b)))$. Using a scaling argument, we can rewrite the surface tension as $\sigma=\int dr(\pd_ r \phi_b)^2$, which allows us to integrate $\sigma$ to obtain
\be 
\s\simeq\lim_{R_b\gg 1 }\frac{  \phi_+^2}{3}\frac{e^{2R_b}(3+e^R_b)}{(1+e^R_b)^3} =  {  \phi_+^2\over3}. 
\ee
The volume contribution from the potential is only a function of the asymmetry  $\Delta V= \frac{1}{2}-\a^2-\frac{2}{3}\a\sqrt{\a^2-2}+\frac{1}{3}(\a^4+\a^3\sqrt{\a^2-2})$. It is convenient to write $\a \simeq 3/2 +\da$, where $\da\ll 1$ in the thin wall approximation. Then, expanding the expressions for the energy, eq. \ref{twe}, and the radius, eq. \ref{twr}, about
$\da=0$, we can obtain approximate analytical expressions for both as follows:

\be 
E_{\rm tw} \approx { C_d \over3} \frac{(d-1)^{d-1}}{2^{2 d-3} d} \left ({1 \over \da^{d-1} }+ {d+3\over \da^{d-2}} \right ) + \mathcal O(\da^{3-d}), \label{dbounce} 
\ee 

 \be 
 R_b \approx {d-1\over 4}    \left ({1 \over \da }+ 1 - 7 \da \right ). 
 \ee  
Although these expressions breakdown quite quickly as $\da$ increases, they provide the dominant scaling of energy and radius with the coupling $\a$.


\subsection{Oscillons: A Brief Review}

Alongside with the critical bubble or bounce, oscillons will play a key role in the mechanism for fast vacuum decay. In fact, the title of this work alludes to their presence in the metastable minimum and their effect on the decay rate. As such, it is useful to briefly review their properties. In their simplest form, oscillons are spatially-extended, very long-lived, time-dependent solutions of the nonlinear Klein-Gordon equation \cite{oscil1}, or of other PDEs with amplitude-dependent nonlinearities \cite{oscgen}. Their properties have been extensively studied during the past decade in two \cite{oscil2d}, three \cite{oscil3d}, and higher \cite{doscil} spatial dimensions, and, more recently, in U(1) Abelian-Higgs models \cite{u1oscil} and in SU(2)xU(1) models \cite{su2oscil}. However, it is fair to say that a more fundamental understanding of their existence and longevity is still lacking.

In the context of relativistic scalar field models, oscillons are characterized by large-amplitude oscillations  about the vacuum state. Assuming spherical symmetry, it has been shown \cite{doscil} that they only exist if these fluctuations probe  beyond the inflection point [$\phi_{\rm inf}$] of $V(\p)$, although they may be more general \cite{MITprivatecom}. The reader may consult the references cited for more details. 

Scalar field oscillons have been found by two methods. The first, and simpler, method makes use of an initial condition $\p(0,r)$ that resembles the oscillon solution, for example, $\p(0,r)=\p_0\exp[-r^2/R^2]$, for $V(\p)$ of eq. \ref{respot}. For values of $R\geq R_{\rm min}$, and $\phi_0 >\phi_{\rm inf}$, the field will evolve into the oscillon configuration, where it will remain for a lifetime that is sensitive to the values of $\p_0,~R_{\rm min}$, and $d$. $R_{\min}$ can be estimated analytically for degenerate and non-degenerate polynomial potentials for an arbitrary number of spatial dimensions \cite{doscil}. 

The second method, more general, shows that oscillons emerge, under very general conditions, when large-amplitude fluctuations about a given vacuum state with sufficient thermal (or quantum noise) are present. In the context of $2d$ models, Gleiser and Howell showed that they emerge after a quench from a single-well to a double-well potential \cite{GH}. In the model of eq. \ref{respot} this would correspond to a change from $\a=0$ to $\a=1.5$.

\begin{figure}
\includegraphics[scale=.6,angle=0]{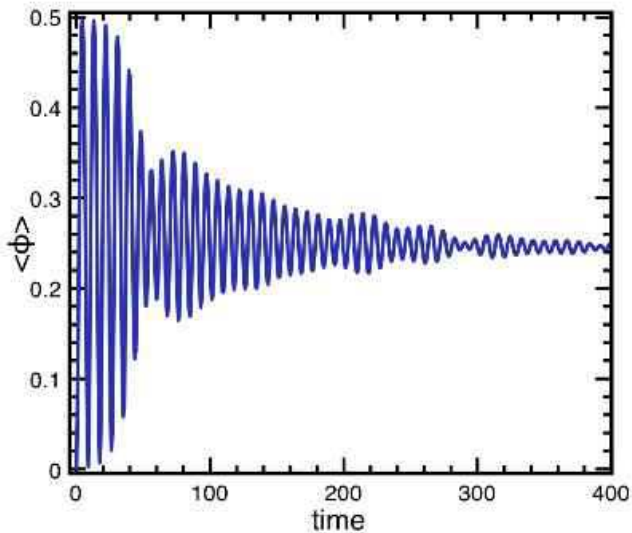}
 \caption{\label{symave}  The volume-averaged field, or order parameter, $\bra \phi(t) \ket \equiv \bar \p(t)$ which corresponds to fig.\ref{symosc}. }
 \end{figure}

Inspecting figure \ref{potential}, one can see qualitatively that the quench induces coherent oscillations of the field's zero mode, ${\bar \p}(t)\equiv (1/V)\int d^dx \p(x,t)$ about the metastable minimum at $\p=0$ due to a ``widening'' of the potential there. Other mechanisms may induce large-amplitude oscillations of the zero-mode, producing similar results. This happens, for example, in the context of reheating in certain models of cosmological inflation \cite{hybrid, resinf}. Another possibility is a symmetry-breaking interaction that changes the shape of the potential from single to double-well, as in the case of the quench above, or one that simply ``widens'' the curvature of the potential about the vacuum. Quantitatively, the energy from these coherent oscillations is transfered to higher $k$ modes via parametric amplification and trigger the emergence of oscillons. We refer the reader to reference  \cite{GH} for details. Not surprisingly, the same behavior ensues in $3d$: a quench induces oscillations of the field's zero mode, as can be seen in figure \ref{symave}. The results were obtained in a cubic lattice of volume $L^3=64^3$ with periodic boundary conditions, lattice spacing $dx=0.5$, and $\a$ changed from $\a=0$ to $\a=1.5$. The initial thermal state was prepared at $T=0.29$. 

\begin{figure*}
\centering
\includegraphics[scale=.7,angle=0]{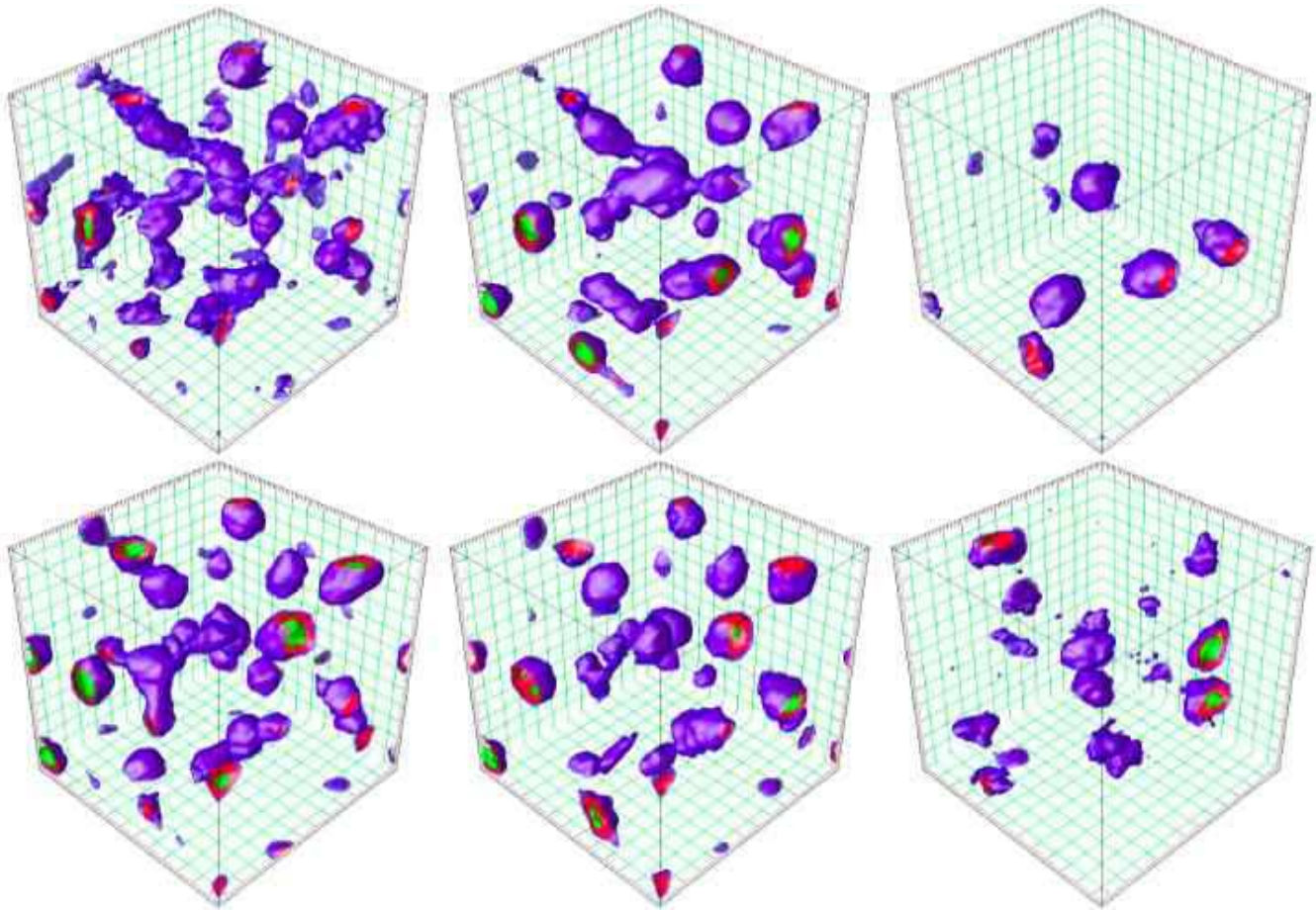}
 \caption{\label{symosc} Snapshots of a $3d$ scalar field simulation after a quench from a single to a symmetric double well potential ($\a=0\rightarrow1.5$). The oscillatory bubble-like configurations which emerge in synchrony are identified as oscillons. The lattice parameters were $dx=0.5$, $T=0.29$ and $L^3=64^3$. Plotted is the $\phi=\{0.75,1.0,1.5\}$ isosurfaces in blue, red and green, respectively.  The plots are snapshots at $t=\{45.5,51.5,54.5,60.5,62.0,194.5\}$, respectively, increasing left to right and top to bottom. Note that oscillons can be tracked over a period of oscillation as they reappear in approximately the same place about $\sim9s$ later. For example, the top right corner of the first, second, fourth and fifth slices show the same oscillon.  The last snapshot shows oscillons still present at late times. They eventually disappear as thermal equilibrium is restored.    } 
\end{figure*}

In figure \ref{symosc}, we show snapshots of the field, displaying the appearance of bubble-like oscillons. Shown are the isosurfaces at $\phi=\{0.75{\rm (blue)},1.0{\rm (red)},1.5{\rm (green)}\}$ [colors online only]. The long-lived, localized oscillations which emerge in synchrony are the oscillon configurations.

\subsection{The Hartree Potential and Lattice Implementation}

As mentioned above, the initial state is prepared as a thermal state centered at $\p=0$ in the potential of eq. \ref{respot} with $\a=0$. The thermal fluctuations will induce changes in the potential for $\p$, which, in leading order in perturbation theory, may be approximated by the Homogeneous Hartree Approximation (HHA). An excellent account of the HHA, using successive truncations within the hierarchy of correlation functions to obtain corrected equations of motion, can be found in ref. \cite{Aarts}.

The HHA assumes that the fluctuations of the field and its associated momentum remain Gaussian throughout its evolution. Thus, the approximation works well just prior and just after the quench for all temperatures, and for low temperatures at all times. The Hartree potential can be derived as follows:  write the field as $\phi= {\bar \phi} +\d\phi$. Then, $ V_H(\phi)= \bra V(\bar{\phi}+\d\phi) \ket$, where $\bra \ket$ means an average over all fluctuations $\d\phi$, with $\bra\d\p\ket=0$ and $\bra \d\p^2 \ket\equiv \b$. $\bra \d\p^2 \ket$ is the translationally-invariant mean-square variance of the field, which, in thermal equilibrium (the initial state), is constant and proportional to $T$. We then obtain (suppressing the bar),
 \be 
 \label{VHartree}
 V_H(\phi)=-\a \b
\phi+
\left(1+\frac{3}{2}\b\right)\frac{\phi^2}{2}-\a\frac{\phi^3}{3}+\frac{\phi^4}{8}.
\ee 

In thermal equilibrium, the momentum and field modes in $k$-space satisfy
\begin{eqnarray}
\langle |\bar{\pi}(k)|^2\rangle &=& T; \\
\langle |\bar{\phi}(k)|^2\rangle &=& \frac{T}{k^2+m_{\rm H}^2},
\label{powerspec}
\end{eqnarray}
where the Hartree mass is defined as $m_{\rm H}^2=V_H''(0)$.

Due to the dependence of  $\langle |\bar{\phi}(k)|^2\rangle$ on the wave number $k$, a lattice implementation of the field theory will introduce a lattice spacing ($\d x)$ dependence \cite{latticespacing}. We thus write, for an UV cutoff $\Lambda$, $\b \equiv a_{\#d}T$, where the constant 
$a_{\#d}$, which changes in different spatial dimensions, can be obtained numerically by comparing the lattice results with eq. \ref{powerspec}. It can also be obtained analytically, as shown in the Appendix. For example, for $\d x =0.2$, we obtain, through this method, $a_{2d}=0.518$ and $a_{3d}=1.194$, in two and three spatial dimensions, respectively. These values agree very well with the numerical estimate. (See Appendix for details.) 

In figure \ref{symphiave}, we plot on the left the ensemble-averaged two-point correlation functions for the field (green squares, bottom) and its related momentum (blue squares, top). The black lines are the analytical formulas of eq. \ref{powerspec}. On the right, we take the logarithm of the initial data set for the two-point correlation function for the field (green squares and black line), and compare to its value during the simulation. The red squares are the data at every half-second for the time interval $360<t<400$, when ${\bar \p}(t)$ is well settled in the potential minimum (check figure \ref{symave}). The blue squares correspond to the same data, but after applying a thermal filter. One can clearly distinguish two populations of modes: low-$k$, oscillon-related modes far from equilibrium (for $k\lesssim 1.5$), and modes that remain in thermal equilibrium for all times ($k\gtrsim 1.5$).

\begin{figure*}
\begin{minipage}{0.5\linewidth}
\centering
\includegraphics[scale=.6,angle=0]{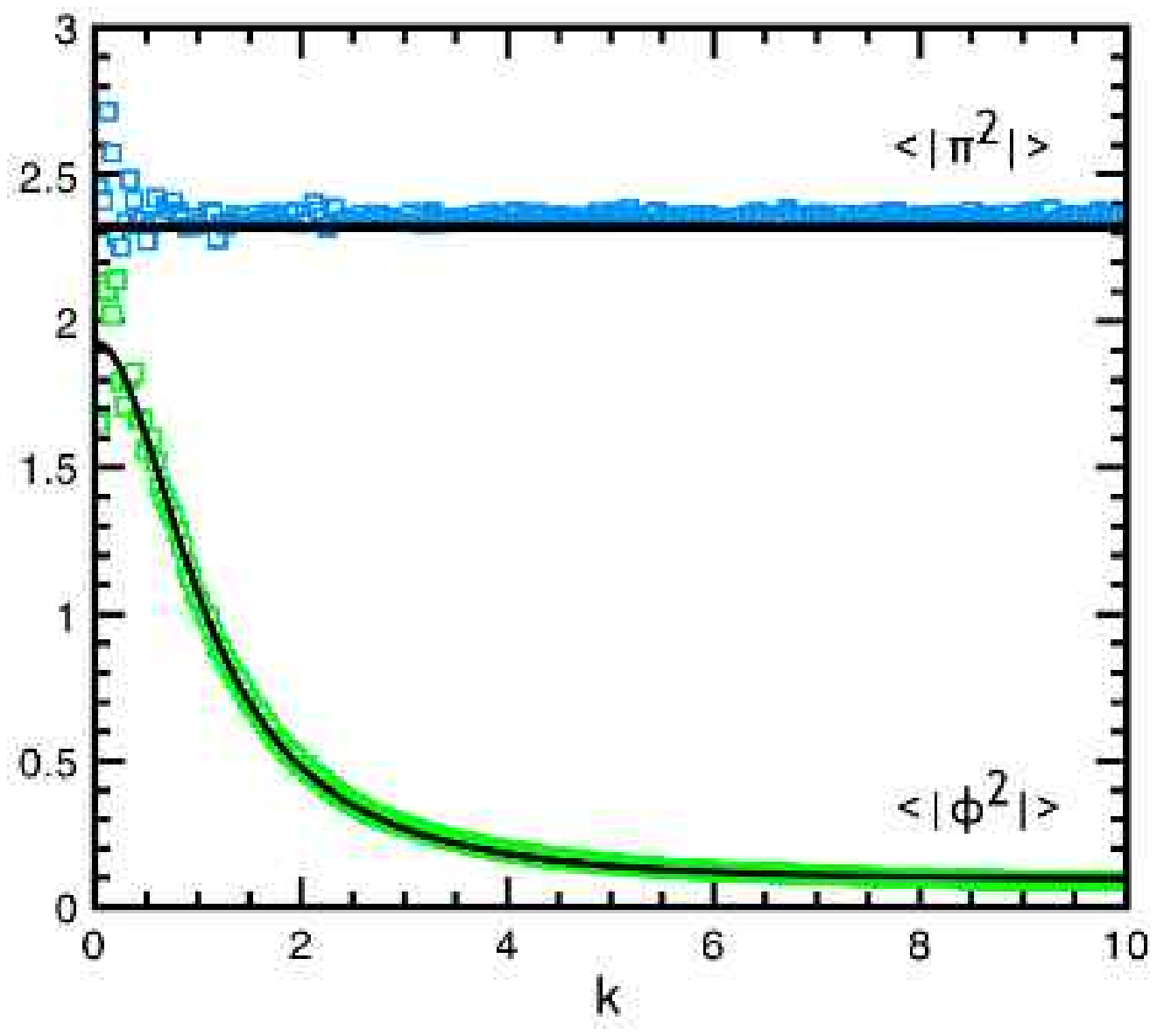}
\end{minipage}%
\begin{minipage}{0.5\linewidth}
\centering
 \includegraphics[scale=.6,angle=0]{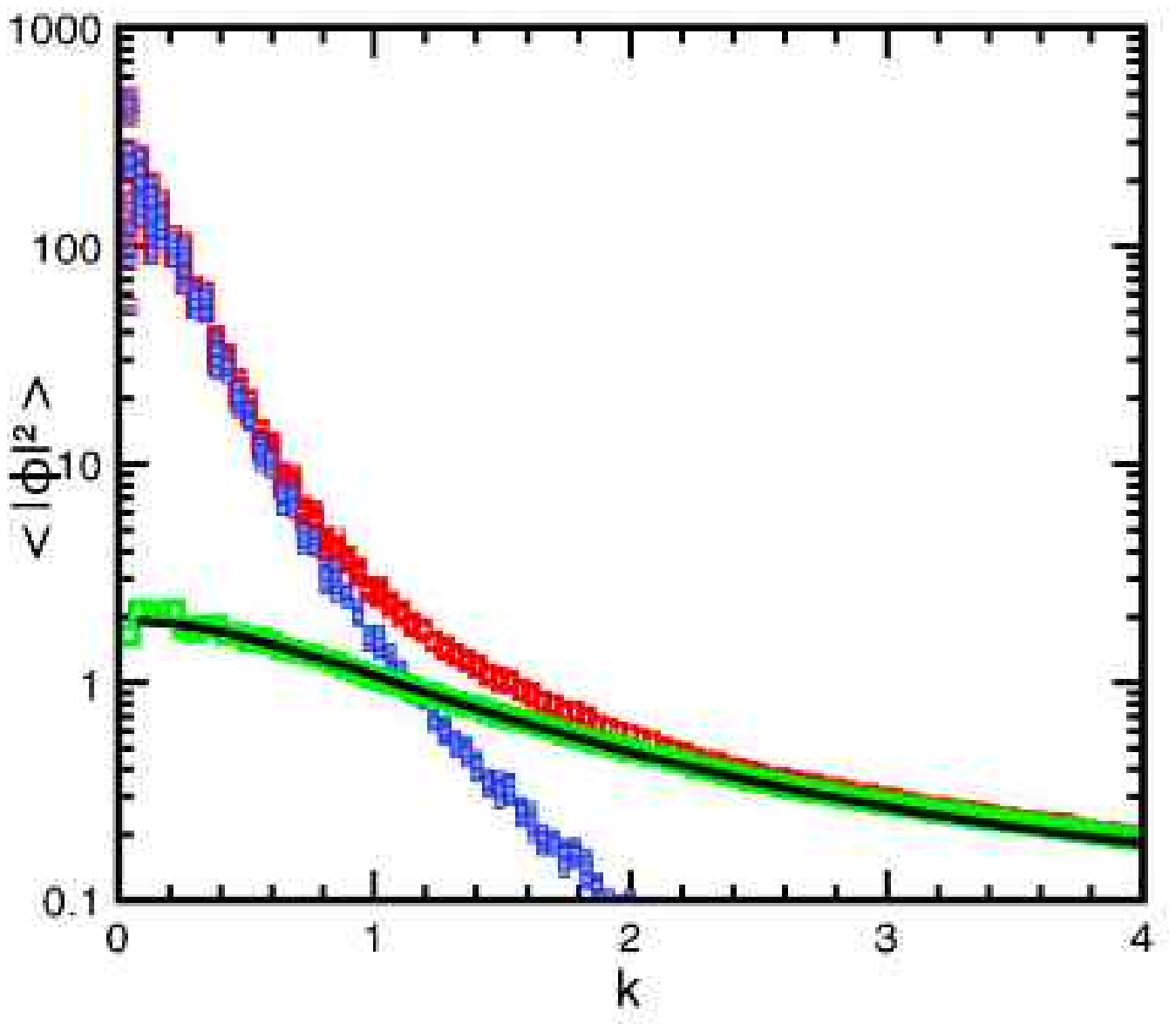}
\end{minipage}
 \caption{\label{symphiave} The radially-averaged 2-point field and momentum correlation functions. On the left, we show the results for the thermal initial conditions. The black continuous lines correspond to $\langle |\bar{\pi}(k)|^2\rangle =T (\d x)^{-3}$ (top) and to  $\langle |\bar{\phi}(k)|^2\rangle = {T \d x^{-3}}/\{k^2+m_{\rm H}^2\} + c_0 \sqrt{k}$ (bottom). (The small constant $c_0=0.024$ used to fit the analytic prediction with the lattice results is due to the abrupt lattice cutoff introduced by $\d x$.) The blue and green squares are from the simulation. On the right, we plot the logarithm of the two-point correlation function for the field. The black line and green squares correspond to the thermal initial state plot on the left. The red squares are the data plotted at every half second for the time interval $360<t<400$. The blue squares are also the data, after applying a thermal filter.  One clearly distinguishes two populations of modes: the oscillon-related, low $k$ modes ($k\lesssim 1.5$), which sharply depart from the thermal spectrum, and the purely thermal modes, which remain in thermal equilibrium throughout the simulation ($k\gtrsim 1.5$). These figures were generated using the data from the same run as in \ fig.\ref{symosc}.}
 \end{figure*}

\begin{figure}
\includegraphics[scale=.6,angle=0]{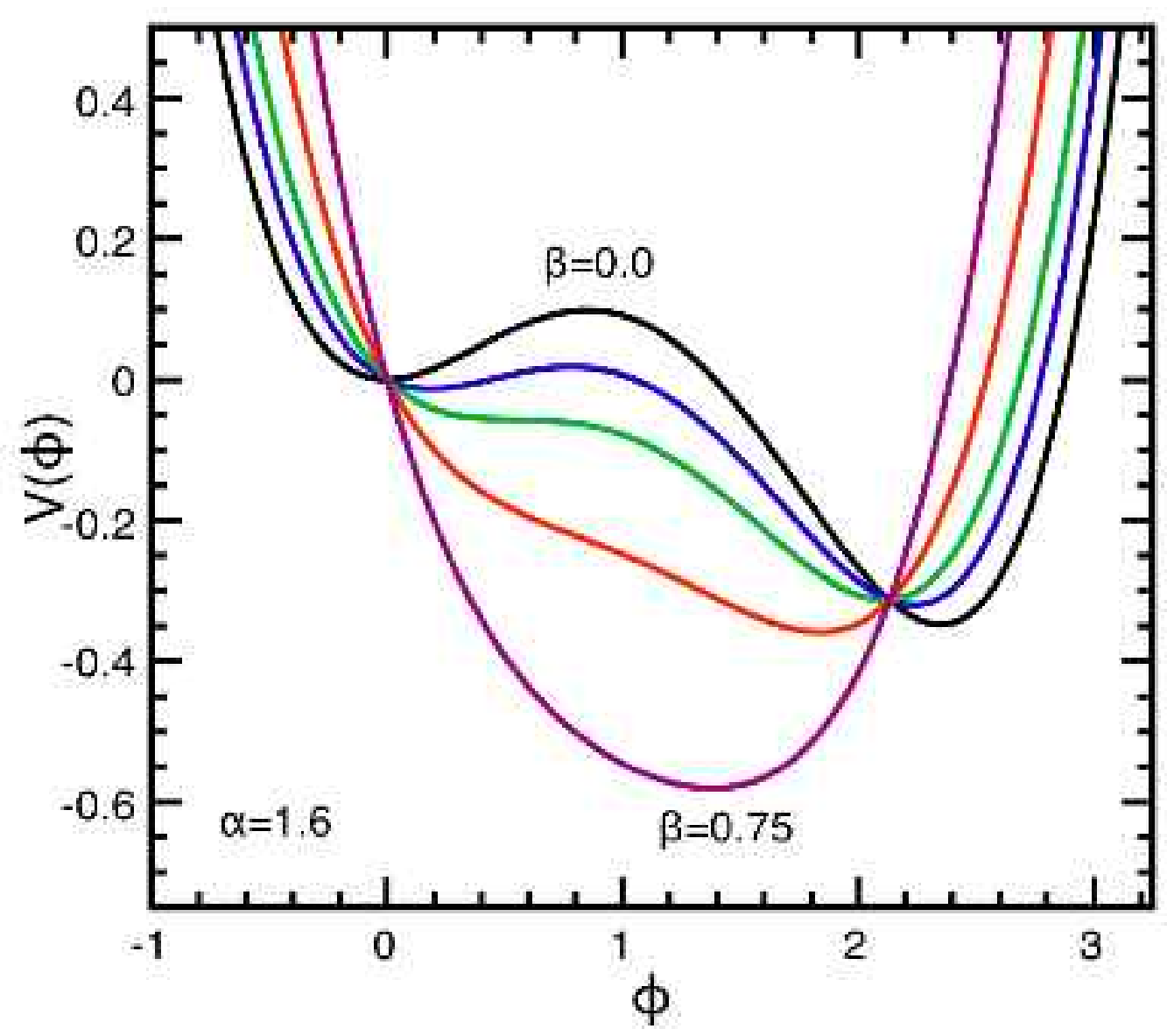}
 \caption{\label{VH} Hartree potential for $\a=1.6$ and several values of $\beta={0,0.1,0.2,0.4,0.75}$.}
 \end{figure}

Introducing $\b$ allows us to relate models with different temperatures and lattice spacings with a single parameter. From now on, we will refer to $\b$ as the {\it effective lattice temperature}.
In figure \ref{VH}, we plot $V_H(\p)$ for various values of $\b$ and $\a=1.6$. The interpretation of $\b$ as an effective temperature should be apparent.

The initial state was prepared by thermalizing the field using standard Langevin techniques \cite{thermalfield}
\be
\label{Langevin}
 \ddot \phi+\g \dot\p -\nabla^2\p= - \pd V/\pd\p + \zeta 
\ee
where $\zeta$ is a Markovian noise with two-point correlation function obeying the fluctuation-dissipation relation,
\be
\bra \zeta({\bf x}, t) \zeta({\bf x}', t') \ket = 2 \gamma T
\delta({\bf x} - {\bf x}') \delta(t - t')~.
\ee
The initial state is solely determined by the temperature of the bath $T$. Note that this thermalization is simply a device to set an initial distribution for the field modes satisfying eq. \ref{powerspec}  . One could have used alternative $T$-independent methods to set up initial conditions. Ours is in tune with the more conventional statement that vacuum decay initiates from a metastable thermal state. This initial thermalization was verified by compliance with equipartition, such that the average kinetic energy of each mode was $T/2$.

We used a standard leapfrog algorithm \cite{NumRec} on cubic lattices with volume $V=L^3=(256*dx)^3$ with periodic boundary conditions. We verified that our results do not show any relevant dependence on $L$. The lattice was evolved in time-steps of $\d t=0.025$. We checked that the Hamiltonian evolution conserves energy to $O(\d t^2)$.

\section{Power-Law Vacuum Decay}

\begin{figure}
\includegraphics[scale=.6,angle=0]{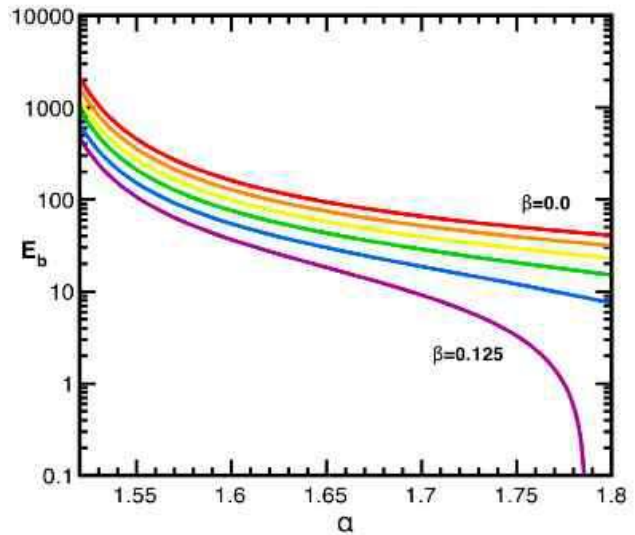}
 \caption{\label{bounceaction} The logarithmic energy of a $3d$ bounce or critical bubble, $\ln[E_{b}(\alpha,\beta)]$, for $\beta=\{0.0,0.025,0.05,0.10,0.125\}$ as a function of $\a$. There is a clear scaling with $\b$ that worsens as $\b$ is increased. For $\b=0$, the bounce ``energy'' is the $d=4$ Euclidean action. } 
\end{figure}

\begin{figure}
\includegraphics[scale=.6,angle=0]{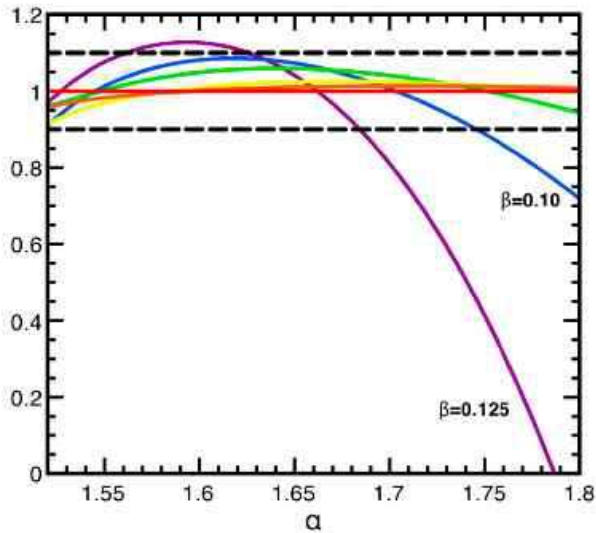}
 \caption{\label{scalingtest} Testing the scaling of the bounce energy with $\b$: we plot the ratio
 $E_b(\a,\b)/\left [E_b(\a,0)A(\b) \right]$ versus $\a$ for various values of $\b$. The scaling holds well (dashed lines denote 10\% accuracy) and worsens as $\b$ increases.} 
\end{figure}

Within the framework of HN, the nucleation energy barrier
can be calculated numerically by integrating the equation: 
\be
\label{bounceeq}
 \pd_r^2 \p +\frac{(d-1)}{r}\pd_r \p=\pd_\phi V_H(\phi,T). 
 \ee
In figure \ref{bounceaction}, we plot the energy barrier $E_b(\a,\b)$ as a function of asymmetry $\a$ for several values of $\b$ for $d=3$. For reference, we also included the Euclidean bounce action for quantum decay at $T=0$. It is apparent that the different curves are related by a $\b$-dependent scaling, 
\be
E_b(\a,\b) \simeq E_b(\a,0)A(\b),
\ee
where we found that $A(\b)=(1-c\a\b)^2$, with $c=2.75$ giving the best fit. In order to test this approximation, in figure \ref{scalingtest} we plot the ratio $E_b(\a,\b)/\left [E_b(\a,0)A(\b) \right]$. It is clear that the scaling holds to within 10\% for the range of parameters where resonant nucleation ensues. The scaling allows us to relate the finite-temperature bounce action computed with the Hartree approximation of eq. \ref {VHartree} with the tree-level bounce action computed with eq. \ref{respot}.

\subsection{Computing the Exponent in Resonant Nucleation}

\begin{figure}
\includegraphics[scale=.6,angle=0]{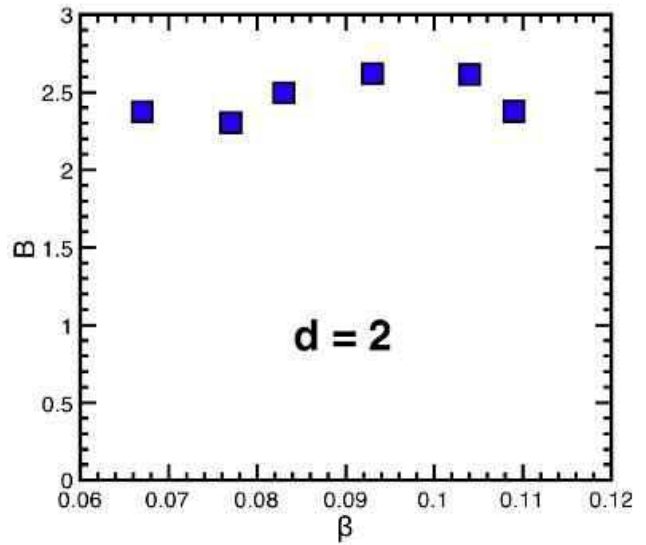}
 \caption{\label{Bslope2d} The value of the decay power $B$ as a function of $\beta$ in two dimensions. } 
\end{figure}

According to HN theory, the decay rate is given by eq. \ref{decayrate}. Is this also true after an abrupt change in the potential $V(\p)$ -- promoted either by a quench, a new interaction, or by a coupling to another field that induces oscillations on ${\bar \p}(t)$, as in certain inflationary models \cite{hybrid, resinf}? In ref. \cite{GH2} (GH), it was found that, in $2d$, changes that induce large-amplitude fluctuations of the field's zero mode may drastically alter the decay rate from an exponential to a power law for a range of asymmetries ($\a$) and temperatures ($T$). In GH, it was argued that the decay is triggered by the resonant nucleation of oscillons, which may coalesce to form a critical bubble or that simply become unstable and grow to become a critical bubble themselves. This mechanism was dubbed {\it resonant nucleation} (RN), and was found to lead to a power-law decay per unit area
\be
\label{RN}
\Gamma \sim T^3\left (E_b/T\right )^{-B},
\ee
with $B \simeq 2.464\pm 0.134$ for the range of parameters where RN is operative. In figure \ref{Bslope2d} we display our results for the exponent $B$ in $d=2$. The lattice spacing was $\d x=0.2$, and the lattice size $2048^2$. Results were averaged over 10 runs. Errors were smaller than boxes.[Note that this value of $B$ and the error bars are smaller than those quoted in \cite{GH2}. Also, we found that $B$ is only weakly dependent on $\b$. This discrepancy is due to better statistics and to a more precise method of distinguishing between RN and cross-over transitions, to be explained below.]

\begin{figure}
\includegraphics[scale=.6,angle=0]{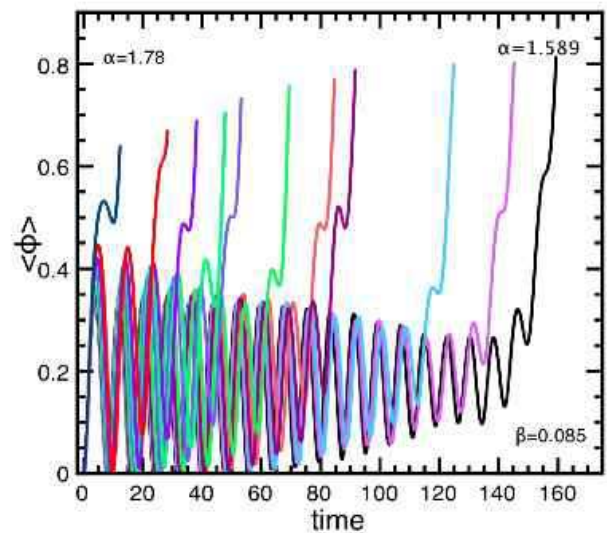}
 \caption{\label{evoplot} The volume-averaged field, or order parameter, $\bar \p(t)$ for $\a=1.589\rightarrow 1.78$ for $\b=0.085$. [Results are better seen in color.]} 
\end{figure}

In order to test if a power-law behavior holds in $3d$, we repeated the GH procedure for different values of the asymmetry $\a$ and the lattice temperature parameter $\b$. The field is initially prepared in a thermal state in a single well ($\a=0$). The potential is then changed to an asymmetric double well, by setting $\a>3/2$. The subsequent dynamics of the field is conservative, that is, we set $\gamma=0$ in eq. \ref{Langevin}. We then measured the volume-averaged value of the field, ${\bar \p}(t)$. Results for $\b=0.085$ and several values of the asymmetry $\a$ are shown in figure \ref{evoplot}.  

One can easily see that, for values of $\a$ close to degeneracy (towards the right of the figure), ${\bar \p}(t)$ performs many damped oscillations about the metastable minimum of $V_H({\bar \p})$ before irreversibly transitioning to the global minimum. For clarity, we cut off the plot before the decay completes. As the asymmetry is increased (towards the left of the figure), the number of oscillations decreases until the transition completes by cross-over, that is, within a time-scale which is shorter than a typical oscillation period. [More on this soon.]

\begin{figure*}
\begin{minipage}{0.5\linewidth}
\centering
\includegraphics[scale=.6,angle=0]{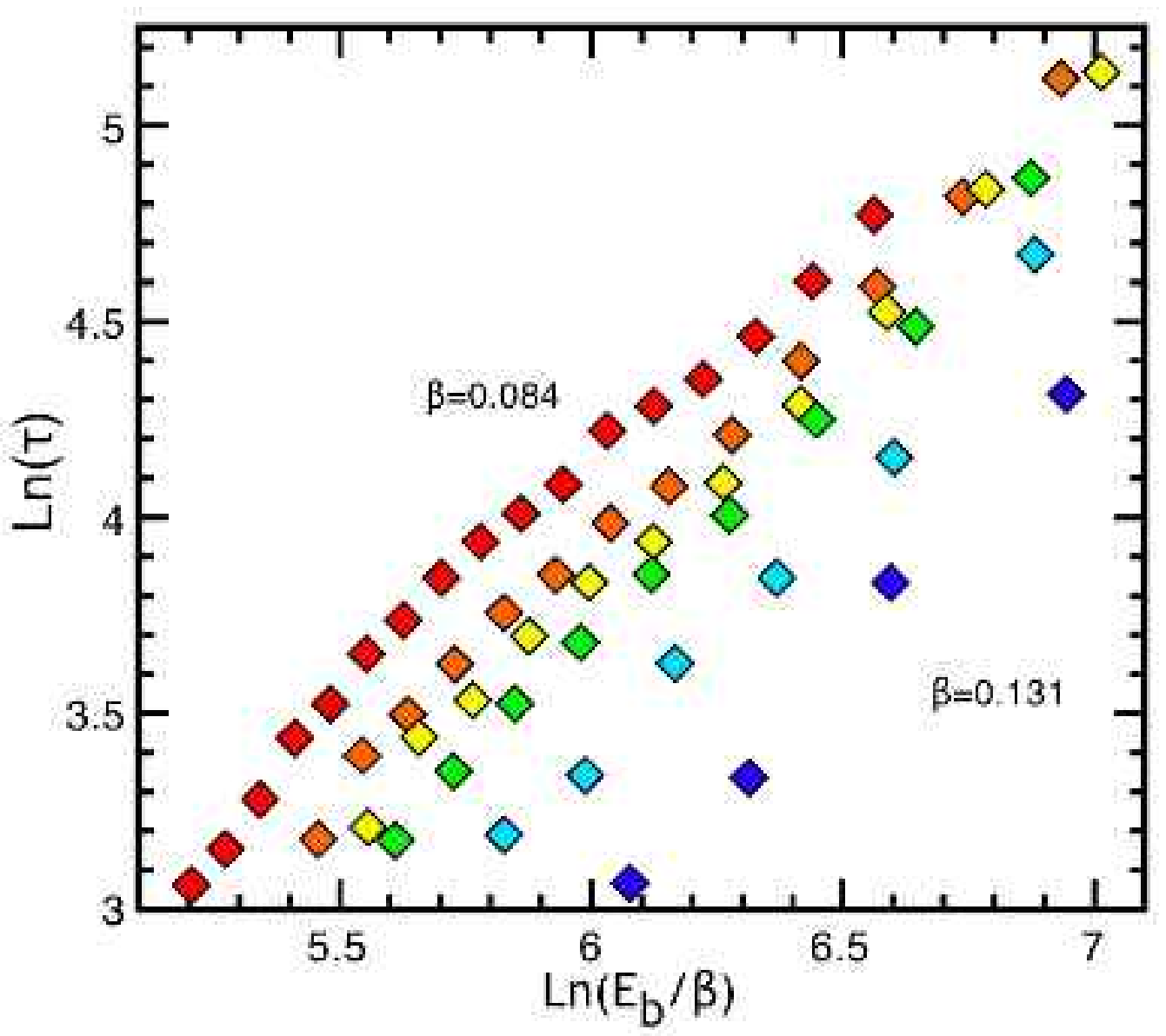}
\end{minipage}%
\begin{minipage}{0.5\linewidth}
\centering
\includegraphics[scale=.6,angle=0]{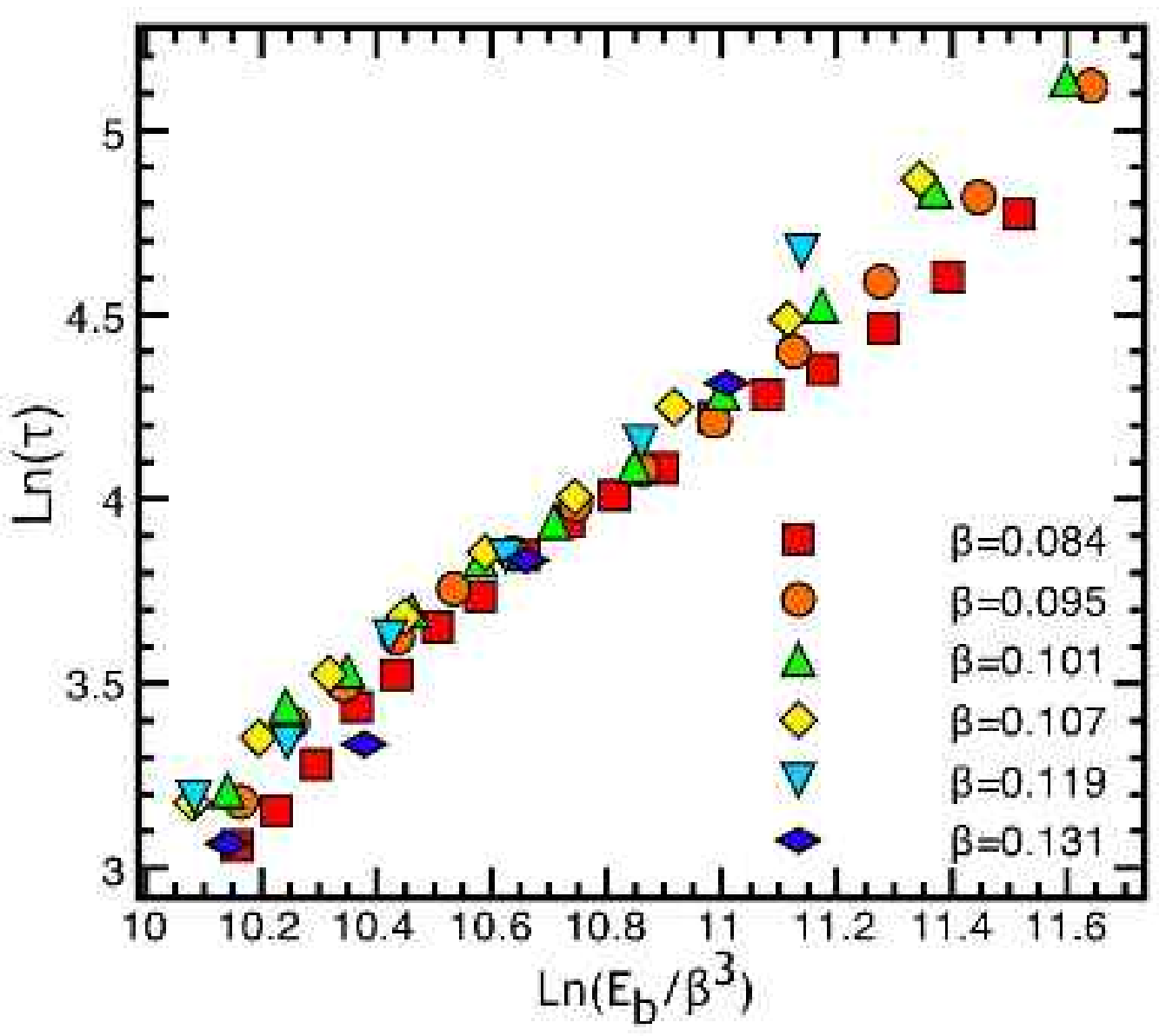}
\end{minipage}
 \caption{\label{decaytime} Decay time-scales in $3d$: we plot $\ln(\tau)$ vs.  $\ln(E_b/\b)$ [left], and $\ln(E_b/\b^3)$ [right]. From top to bottom on the left figure, $\b=\{0.084, 0.095,0.101,0.107,0.119,0.131\}$. }
\end{figure*}

Adopting as the decay time-scale ($\tau$) the time at which ${\bar \p}(t)$ crosses the maximum of $V_H({\bar \p})$, we can plot the logarithm of decay times as a function of the logarithm of the bounce energy $(E_b/\b)$, where $E_b$ is computed from the solution of eq. \ref{bounceeq} for a given pair ($\a,\b$). The results are ensemble-averaged over 50 runs, and displayed as individual points (diamonds) in the left plot of figure \ref{decaytime}. On the right side, we plot the  logarithm of $(E_b/\b^3)$ to show the simple scaling with $\b$.
For reference, if $\delta x=0.25$, the value we adopted in our simulations, the temperature would be in the range $0.09\leq T\leq 0.14$. [To get the value of the temperature, use that $T=\b/a_{3d}$. Analytically, $a_{3d}=0.955$ (see Appendix), and numerically $a_{3d}=0.933$, an error of only 2.4\%.] The slopes of the straight-line fits are the numerical values for the decay power $B$. 

\begin{figure}
\includegraphics[scale=.6,angle=0]{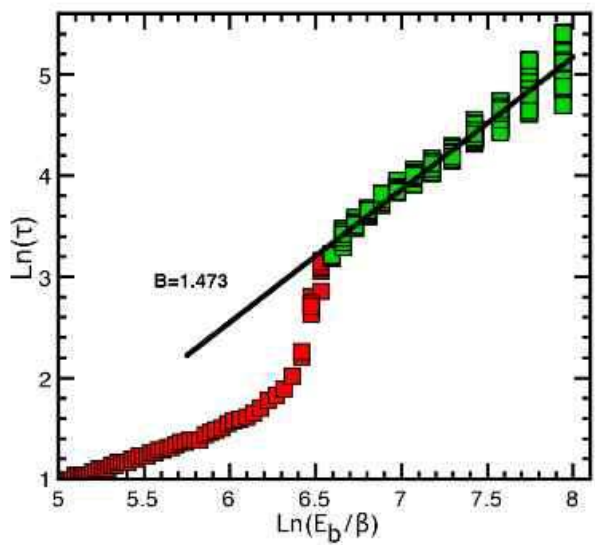}
 \caption{\label{b0954} Extracting the decay exponent $B$: shown are the raw data points for the decay time $\ln\tau$ vs. $(\ln E_b/\b)$ with
 $\b=0.0954$. We ran 50 experiments for each value of $\a$ (or, equivalently, of $E_b$). In this illustration, we used the tree-level potential of eq. \ref{respot} to compute $E_b$. The straight-line slope fit gives $B=1.473$. Note the clear departure from a straight-line fit for $\ln\tau\lesssim 3.2$. We characterize time-scales faster than those as typical of cross-over transitions.} 
\end{figure}

Note that we are not including decay times smaller than $\tau \simeq 22$. Times shorter than this are characteristic of a cross-over transition and would not fall under resonant nucleation. This can be made clear by investigating how we extracted the slopes of the log-log plot, that is, the constant $B$ controlling resonant nucleation. As an illustration, fixing  $\b=0.0954$, consider the results for the decay times displayed in figure \ref{b0954}. From the data, it is clear that there is a sharp departure from a straight-line fit below $\ln(\tau)\simeq 3.2$. This happens consistently for all parameters we investigated, offering a natural cutoff below which the transition occurs via cross-over.

\begin{figure}
\includegraphics[scale=.6,angle=0]{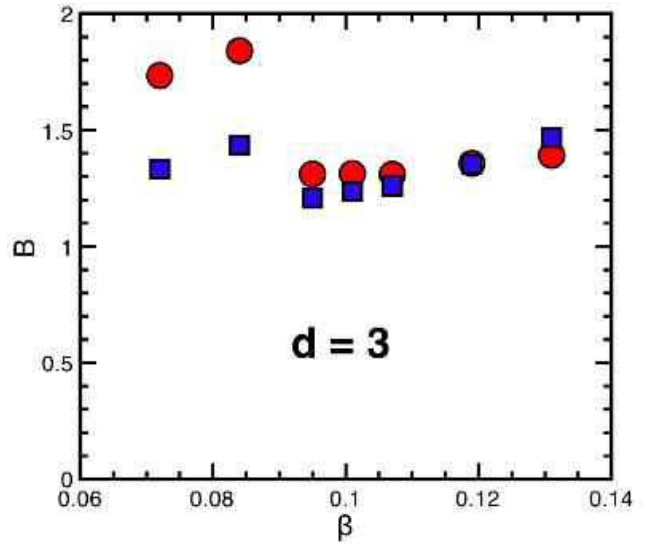}
 \caption{\label{Bslope}  Results for the decay power $B$. Slopes as measured individually, as in figure \ref{b0954}. The circles are calculated with the tree-level potential and the squares with the Hartree effective potential. Error bars from ensemble-averaging fit within the sizes of the squares and circles.} 
\end{figure}

In figure \ref{Bslope}, we show our results for the power $B$ controlling resonant decay. We have included two data sets: the squares are the results obtained using the Hartree effective potential to compute the bounce action and to read off the nucleation time-scale (which depends on the location of the maximum). The circles are the results obtained using the tree-level potential. It is clear that for $V_H(\p)$, $B$ is fairly independent of $\b$ within the range where resonant nucleation applies. The average value is $B=1.327\pm0.059$. The values agree well for $\b>0.09$. For smaller $\b$, it is necessary to increase the value of $\a$ to probe into the resonance nucleation regime, leading to larger errors in the tree-level estimates. 

We also verified that the value of $B$ is not very sensitive to changes in lattice spacing. As an illustration, let us choose $\delta x=1$. For $T=0.55$, which
gives $\beta=0.131$, we repeated the numerical experiments to obtain $B\simeq 1.68$. By adjusting $T$, this value can, for example, be  compared with the one obtained with lattice spacing $\d
x=0.25$ for the same $\beta=0.131$, which we measured to be $B=1.47$. This is a change of only 13\% in $B$ for a factor $4$  change in lattice spacing. 
As a second illustration, we compared the results for $\delta x=0.35$ and $T=0.175$, for which $\beta=0.119$, with $\delta x=0.25$, at same $\beta$.  The values for the power-law decay were $B=1.458$ and $B=1.353$ respectively, a $7\% $ 
difference. Within the accuracy of our simulations, we conclude that $B$ is very weakly dependent on $\d x$ and, most importantly, on $\b$.

\subsection{Phase Diagram for Vacuum Decay}

We summarize our results  in a phase diagram displaying the three possible mechanics for false vacuum decay as a function of the parameters controlling the asymmetry of the potential ($\a$) and the temperature of the initial metastable state ($\b$). Homogeneous nucleation (dark blue region labeled as HN) describes vacuum decay for low enough temperatures for all values of asymmetries. 

Resonant nucleation (RN) falls within the region bracketed by the two left to right curves (light blue region labeled as RN). The diamonds and circles are the results of the simulations. The circles, being close to the cross-over region ($\tau \sim 25$), were not used to compute the decay power $B$. The curves bracketing RN from above (cross over, purple region) and below (HN region) were obtained by using the average value of the field, ${\bar \phi}(t)$, as a guideline: a field well-localized within the false vacuum (${\bar \phi}(t) \ll \phi_{\rm inf}$) will only decay via HN; a field for which ${\bar \phi}(t)\gg  \phi_{\rm inf}$ will easily cross over. RN lies in between. Noting that the amplitude of thermal fluctuations about ${\bar \phi}(t)$ is given by $|\sqrt{\langle\delta\phi^2\rangle}| = |\sqrt{\beta/2}|$, we obtain the condition,
\be
\label{RNregion}
V_H(\p_{\rm inf} \pm \sqrt{\beta/2})= 0.
\ee
This condition translates into a relation between $\a$ and $\b$ that generates the two curves bracketing the RN region horizontally. 

\begin{figure}
\includegraphics[scale=.6,angle=0]{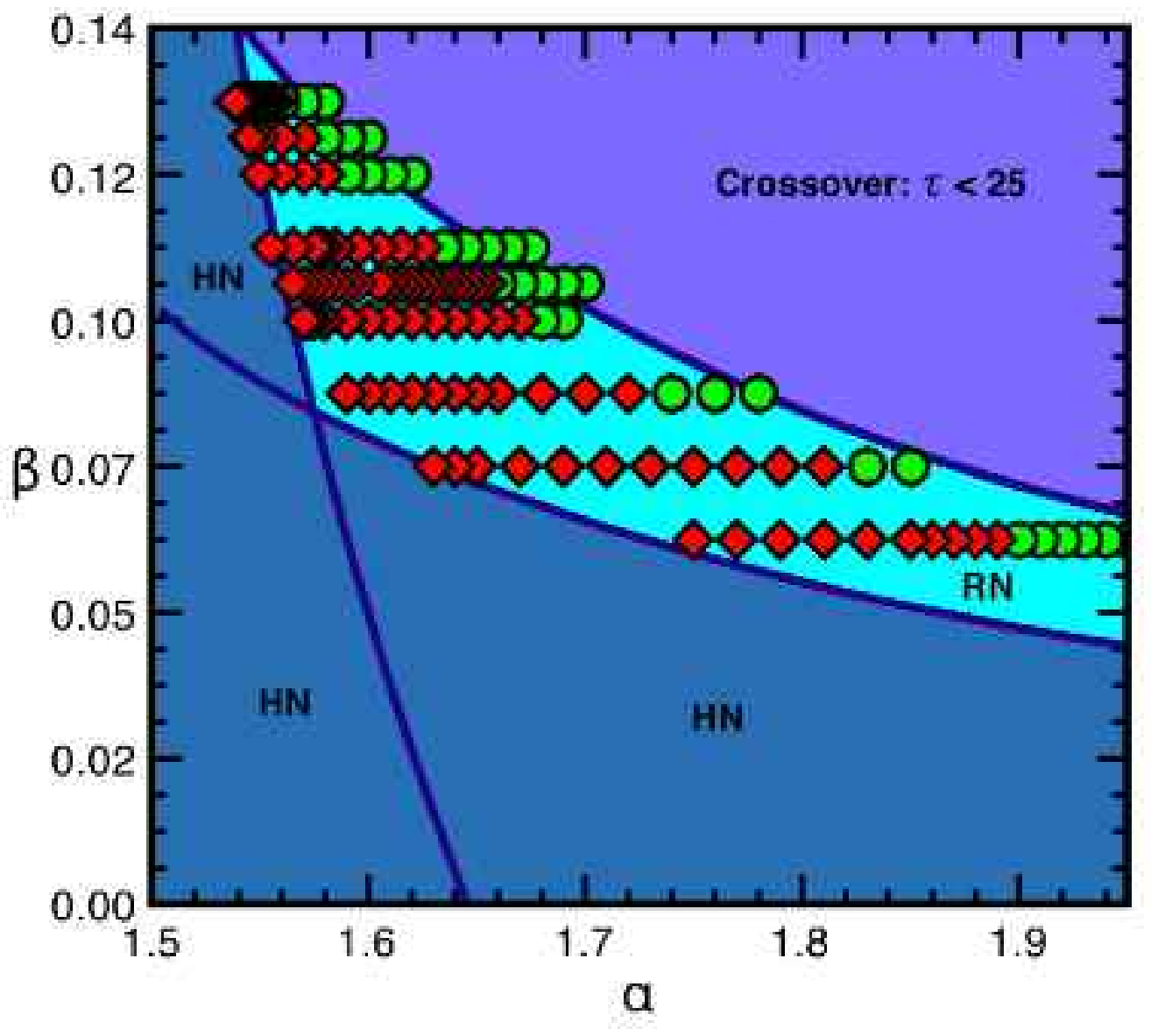}
 \caption{\label{phaseplot} Phase diagram for false vacuum decay. Shown are the three regions characterizing vacuum decay as a function of the asymmetry in the Hartree potential $\a$ and the effective lattice temperature $\b$. HN (dark blue) represents decay by homogeneous nucleation of critical bubbles. RN (light blue) represents decay by resonant nucleation, either by the coalescence of two oscillons (diamonds) or by the unstable growth of a single oscillon (circles). The top region represents fast cross-over, within time-scales of $\tau < 25$. [The regions were labeled for clarity with B\&W display.}
\end{figure}

\begin{figure}
\includegraphics[scale=.6,angle=0]{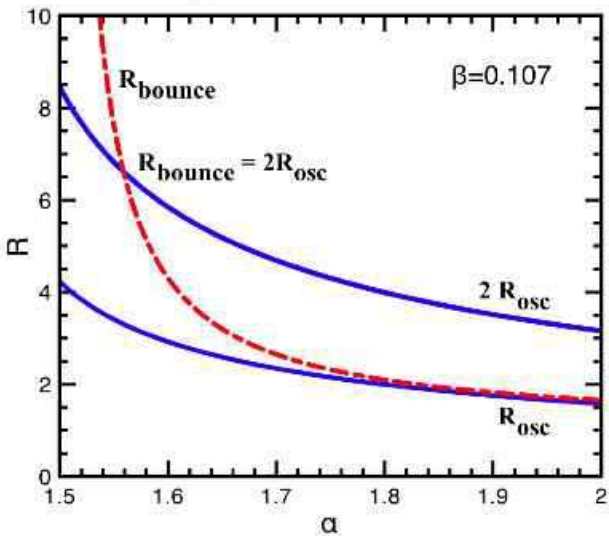}
 \caption{\label{OscBounce} Comparing the sizes of oscillons [$R_{\rm osc}$ -- continuous (blue) lines] and critical bubbles [$R_{\rm bounce}$ -- dashed (red) line] as a function of asymmetry $\a$: $2R_{\rm osc}$ is the upper line, and $R_{\rm osc}$ lower continuous (blue) line. 
Data obtained for $\beta=0.107$ . }
 \end{figure}

The steeper line to the left is obtained by examining the dynamics of resonant nucleation. As noted in ref. \cite{GH2}, in the region near degeneracy, where the critical bounce is large, RN will occur due to the coalescence of two or more oscillons. In figure \ref{OscBounce}, we compare the radius of a bounce with that of oscillons as a function of $\a$ for $\b=0.107$. There are three distinct regions: For large $\a\gtrsim 1.7$, the bounce and oscillon practically match. This region is typical of quick cross-overs or at most borderline RN, where a single oscillon becomes unstable, grows, and becomes critical. In the intermediate region, where $1.56 \lesssim \a\lesssim 1.7$, $R_{\rm bounce} < 2R_{\rm osc}$: here, two oscillons may coalesce to form a critical bubble. This is the region typical of RN. Lastly, for $\a<1.56$,  $R_{\rm bounce} > 2R_{\rm osc}$ and more than two oscillons are needed to produce a critical bubble, a process that, although not impossible, is strongly suppressed in the allowed time scales where RN can act, that is, while oscillons are present in the system, typically while $\partial_t\bar\phi(t)\neq 0$. In this region, HN is the favored decay route. The steeper line in the phase diagram is thus obtained by tracing the points where $R_{\rm bounce} = 2R_{\rm osc}$ for each $\b$, in excellent agreement with the simulations: RN, denoted by the diamonds, ends on this line.

\begin{figure*}
\includegraphics[scale=0.7,angle=0]{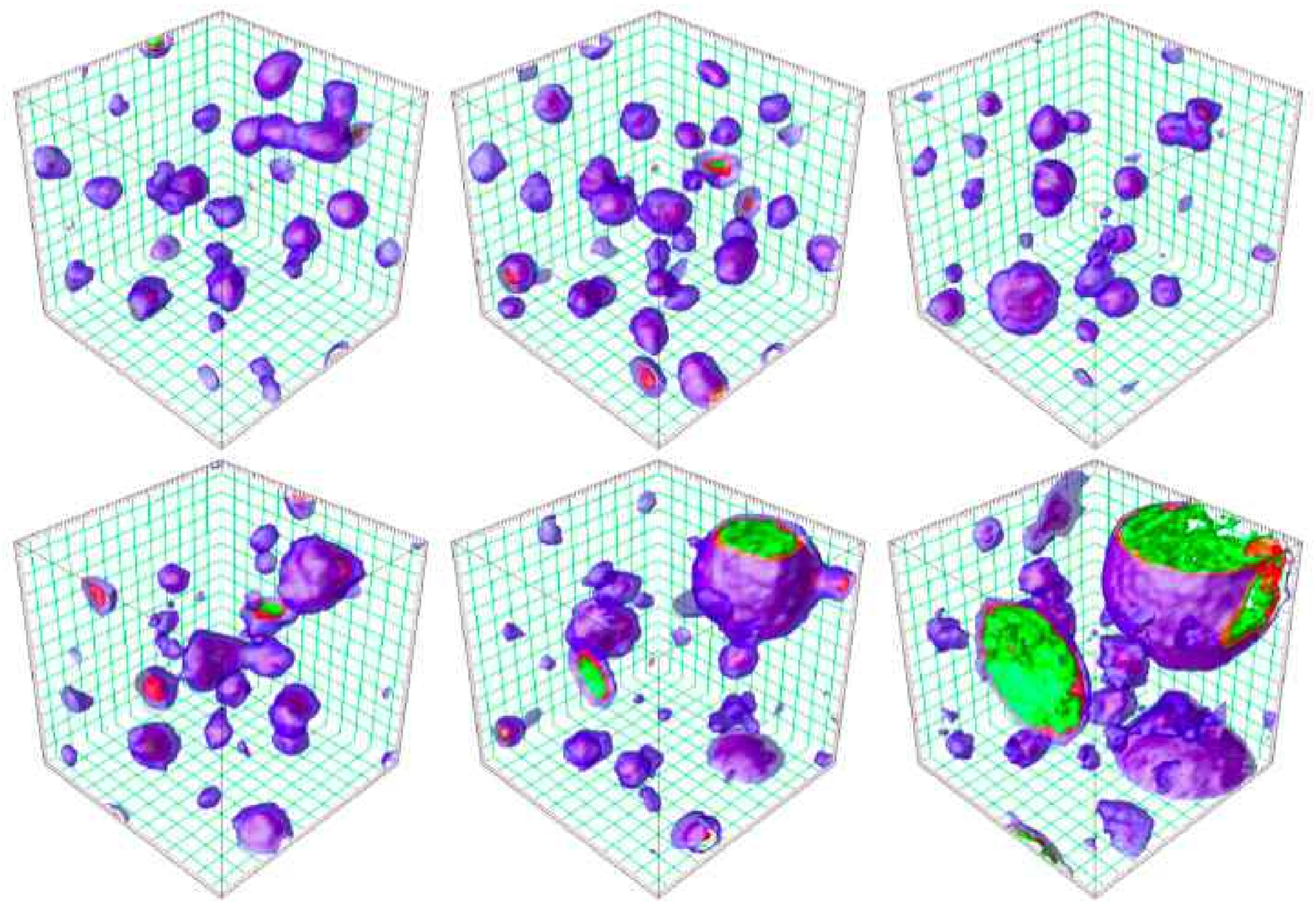}
 \caption{\label{oscilemerge} Snapshots of a $3d$ scalar field simulation after a quench from a single to an asymmetric double well potential ($\a=1.545$, $T=0.28$, $dx=0.5$, $\b=0.134$). The oscillons coalesce to become a critical bubble that grows to complete the vacuum decay. The slices are at times $t=\{27.0,38.0,45.5,54.5,71.0,79.5\}$. The blue, red and green isosurfaces are at $\phi=\{0.9,1.75,1.85\}$, respectively. The average value of the field along with the respective isosurface locations are also noted on fig.\ref{evoplotasym}.  } 
\end{figure*}

\begin{figure}
\includegraphics[scale=.6,angle=0]{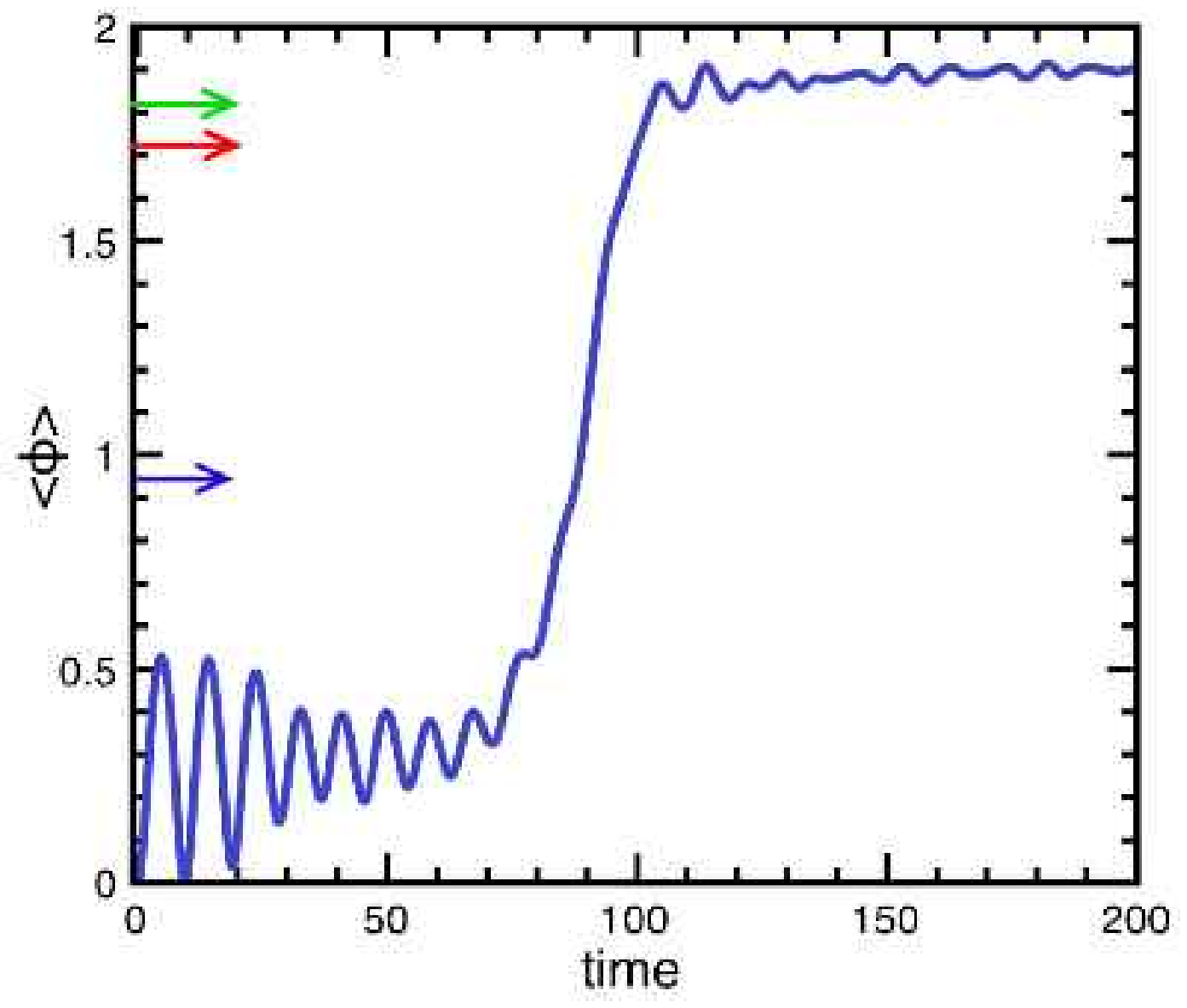}
 \caption{\label{evoplotasym} The volume-averaged field, or order parameter, $\bra \phi(t) \ket \equiv \bar \p(t)$ for $\a=1.545$ for $\b=0.1385$. The colored arrows label the locations of the isosurfaces depicted in figure \ref{oscilemerge}. }
 \end{figure}

In figure \ref{oscilemerge}, we show the coalescence of oscillons leading to a critical bubble that grows to complete the vacuum decay. This figure should be contrasted with figure \ref{symosc} for a symmetric double-well potential. The regions displayed are isosurfaces at $\p=\{0.9,1.75,1.85\}$ which are colored blue, red, and green, respectively. We used $\a=1.545$ and $\b=0.134$. In figure \ref{evoplotasym} we show the corresponding evolution of the volume-averaged field, ${\bar \p}(t)$. The arrows denote the locations of the isocurvatures depicted in figure \ref{oscilemerge}.

\section{Summary and Outlook}

We have presented a detailed numerical study of the decay of metastable vacua in scalar field theories. In particular, we investigated the effects of large-amplitude fluctuations about the vacuum state on the decay rate. These large-amplitude fluctuations were produced by a rapid quench from a single to an asymmetric double well potential. We have argued that they are characterized by spatially-extended, long-lived oscillatory configurations called oscillons. There are, of course, short-lived, small-amplitude configurations as well, but those play a small role on the dynamics of vacuum decay. They are incorporated, via a perturbative approach, into the effective Hartree potential. 
	
	Our main result was to obtain a phase diagram characterizing the three possible mechanisms for vacuum decay. Two of them, slow homogeneous nucleation and rapid cross-over, are well known. Homogeneous nucleation ensues when fluctuations about the metastable vacuum are small-amplitude, within the perturbative regime. Cross-over ensues when fluctuations are very large-amplitude, probing well beyond the inflection point of the effective potential. The novel mechanism, resonant nucleation, lies between these two and ensues when fluctuations probe the neighborhood of the inflection point. We have shown that, as in $2d$ \cite{GH2}, $3d$ resonant nucleation is characterized by a power-law decay rate, $\tau\sim (E_b/\b)^B$, where $E_b$ is the energy of the critical bubble computed with the effective potential, $\b$ is the effective lattice temperature, and $B=1.327\pm 0.059$ is the exponent controlling the decay. We argued that resonant nucleation typically occurs when two oscillons coalesce to form a critical bubble or, marginally, when a single oscillon becomes unstable to growth. Although the coalescence of three or more oscillons is possible, this process tends to be highly suppressed due to its long time-scale.
	
	Our results open many avenues for future investigation. Can oscillons play a role in inflationary cosmology? During reheating, the inflaton undergoes damped, large-amplitude oscillations about the potential minimum. This is essentially the same mechanism that gives rise to oscillons. Of course, one must add the effects of the cosmic expansion, but preliminary results in simple $1d$ models indicate that oscillons not only will be present but will actually be stabilized by the expansion, becoming spikes in the energy-density distribution \cite{GrahamStamatos}. We are currently investigating such possibility in a full $3d$ simulation. Another area where the effects of oscillons should be further investigated is during spontaneous symmetry breaking. In the present work, we investigated a simple $Z_2$ symmetry breaking, as we quenched from a single to an asymmetric double well potential. It is clear that the dynamics of symmetry breaking is very sensitive to the model parameters and that oscillons can play a key role. In a related work, we found that low-momentum vortex-antivortex scattering in broken U(1) gauge models can lead to very long-lived oscillons characterized by a (practically) non-radiating oscillating magnetic dipole and nontrivial Chern-Simons number \cite{u1oscil}. Farhi {\it et al.} and Graham \cite{su2oscil} found a long-lived oscillon in broken SU(2)xU(1) models when the Higgs mass is twice the $W^{\pm}$ mass. If this turns out to be the right mass ratio, as hopefully we will soon know from the LHC, we should expect electroweak oscillons to exist in Nature. Alternatively, they may exist even for a wider range of values. Taken together, these results demonstrate the rich physics of time-dependent, spatially-extended field configurations. This richness is only beginning to be explored. 

\acknowledgments
MG and JT were partially supported by a National Science Foundation grant PHY-0653341. We also would like to thank the Discovery parallel network at Dartmouth for access to their facilities and the NCSA Teragrid cluster for access under grant number PHY-070021. We would also like to thank Massimo di Pierro for help getting our parallel codes running with FERMIQCD \cite{Di Pierro:2003sz}.

\section{Appendix: Analytical Estimate of Effective Lattice Temperature}

Consider the simple case of a quadratic potential $V_0=\frac{1}{2}\p^2$, for which the Hartree potential is simple $V_H={1\over 2} \p^2 + {1\over 2}\beta $.  Consider also the 1-loop potential with UV cutoff $\Lambda$,
\be
V_{\rm 1L}(\phi)=V_0 + {T \over 2}\int_0^{\Lambda}{{d^dp}\over
{(2\pi)^d}}{\rm ln}\left (p^2 + V_0''\right ).
\ee
In $d=2$ the integral gives,
\be
V_{\rm 1L}(\phi)= V_0 + \frac{T}{8 \pi} V''_0 \left \{1 - \ln
\left(\frac{V''_0}{\Lambda^{2}}\right)\right\},
\ee
while in $d=3$ we obtain
\be
V_{\rm 1L}(\phi)= V_0 + \frac{T}{12 \pi^2} \left\{ 3 V''_0 \Lambda - \pi (V''_0)^{3/2} \right\}.
\ee
Focusing on the simple quadratic potential, for which $V_0''=1$, we now match the cutoff-dependent term of $V_{\rm 1L}$ (multiplied by $d/2$) with the $\beta$-dependent term of $V_H$ (eq. \ref{VHartree})  to obtain,
\be 
\b=\frac{T}{4 \pi}\left (1+\ln\left(\frac{\pi^2}{\d x^2}\right)\right)\equiv a_{2d}T,~[2d]
\ee
and
\be 
\b= \frac{3 T}{4\pi \d x}\equiv a_{3d}T,~[3d]
\ee 
where we used $\Lambda=\pi/(\d x)$ and introduced, for convenience,
the lattice-dependent constant $a_{\#d}$. For example, for $\d x =0.2$, $a_{2d}=0.518$ and $a_{3d}=1.194$. This value agrees very well with the numerical estimate, obtained by comparison with eq. \ref{powerspec}.  

After comparing terms, one can write $a_{\#d}$ most generally as \be a_{\#d} \hat{\sim}   \frac{d}{2} \int_0^\Lambda \frac{d^d p}{(2\pi^d)} \ln(1+p^{-2}) \label{alattd} \ee where $\hat{\sim}$ implies take the first term of the Taylor series expansion of the integral should be $a_{\# d}$.   In the Table below, we compare the numerical and analytical (eqn. \ref{alattd}) measurements of $a_{\#d}$ in different dimensions.

\vspace{0.25cm}

\centering
\begin{tabular}{|c|c|c|c|} \hline
$a_{\#d}$ & {\rm analytical} & {\rm numerical} & {\rm lattice size} \\ \hline
$a_{1d}$ & 0.5                      & $0.5\pm 0.001$     & $10^6$ \\ \hline
$a_{2d}$ & 0.518                 & $0.53\pm 0.005$   & $1024^2$ \\ \hline
$a_{3d}$ & 1.194                 & $1.19\pm 0.005$   & $ 100^3$ \\ \hline
$a_{4d}$ & 3.125                 & $3.84 \pm 0.005$  & $32^4$ \\ \hline
\end{tabular}

\vspace{0.25cm}

With this simple analytical method, it is possible to relate the initial state temperature $T$ to the effective lattice temperature $\b$ for any choice of lattice spacing.


\end{document}